\documentclass[12pt]{iopart}
\usepackage{graphicx}

\begin{document}

\title[]{Propagation of strangelets in the Earth's atmosphere}

\author{Fei Wu$^1$, Ren-Xin Xu$^2$, Bo-Qiang Ma$^{3,1}$}
\address{$^1$Department of Physics, Peking University, Beijing 100871, China}
\address{$^2$Department of Astronomy, Peking University, Beijing 100871, China}
\address{$^3$CCAST (World Laboratory), P.O.~Box 8730, Beijing 100080,
China}

\ead{r.x.xu@pku.edu.cn}

\begin{abstract}
A new model for the description of strangelets' behavior in the
Earth's atmosphere is presented. Strangelet fission induced by
colliding with air nuclei is included. It is shown that
strangelets with certain parameters of initial mass and energy may
reach depths near the sea level, which can be examined by
ground-based experiments.
\end{abstract}
 \submitto{\JPG}

\section{Introduction}
\label{}

In a seminal work about two decades ago, Witten proposed that
strange quark matter (SQM), the combination of roughly equal
number of up, down and strange quarks, might be the true ground
state of quantum chromodynamics (QCD)~\cite{witten1984}.
Later calculations have shown that in spite of the effect of their
finite volume, small nuggets of SQM in form of ``strangelets'' can
also be stable as long as the baryon number exceeds a critical
value $A_{crit}$~\cite{farhi1984}.
On the one hand, various theoretical scenarios have provided
chances for strangelet formation~\cite{madsen2003,zhang2005}. They
could be produced in highly energetic nuclear
collisions~\cite{greiner1987}, might originate from the collisions
of two strange stars~\cite{madsen1988}, and could also be ejected
by supernova explosions~\cite{vucetich1998}.
On the other hand, several exotic cosmic ray events have been
reported by balloon and mountain
experiments~\cite{kasuya1993,ichimura1993,price1978}, which are
considered to be ideal candidates of strangelets.
The ultra-high energy cosmic ray (with energy $>
10^{19}~\textrm{eV}$ ) events could also be the results of
extensive air showers of relativistic strangelets accelerated in
pulsar magnetospheres ~\cite{madsen2003,xw03}.
Interestingly, one doubly charged event, with charge-mass ratio of
$\sim 0.1$, has been detected by the AMS experiment in space
~\cite{co03} and suggested to be strangelet-originated. This idea
could be tested in the future AMS02.

Since the existence of stable SQM would have remarkable
consequence in cosmology and astrophysics, what is experimentally
important is to find out strangelets' contribution to cosmic ray
flux and the mechanism for the propagation of strangelets in the
Earth's atmosphere, both of which are helpful to confirm their
existence.
Recently, Madsen has estimated the flux of strangelets in cosmic
rays incident on the Earth~\cite{madsen2005}.
As for the latter, unfortunately, the necessary interaction input
physics is at best poorly known.

Different phenomenological models of strangelet penetration in the
atmosphere have been used  by several authors to explain exotic
cosmic ray events.
Wilk {\it et al.} conjectured that although the initial mass of
strangelets might be quite large, it decreases rapidly due to
collisions with air molecules, until the mass reaches a critical
value below which the strangelet disintegrates into
neutrons~\cite{wilk1996}.
Banerjee {\it et al.} provided a quite different scenario in which
a strangelet picks up mass from atmospheric
atoms~\cite{banerjee2000}.
Monreal novelly discussed the issue of strangelet accumulation in
the atmosphere~\cite{monreal2005}.

In the present work, we will reinvestigate this issue within the
framework of rotating liquid drop model, which is still a
phenomenological model.
We assume that SQM nuggets produced from any cosmological or
astrophysical objects do reach the surface of the atmosphere, and
we evaluate their behavior in the atmosphere.
We find that strangelets with particular initial baryon number and
particular initial Lorentz factor can reach mountain altitudes,
even the sea level($\sim 1000~\mathrm{g~cm^{-2}}$).

In the following sections, we will first provide revised results
of ground state strangelet calculated from liquid drop model.
Then we will investigate the colliding cross section between a
liquid strangelet and an air nucleus.
Finally we will give numerical results about the propagation of
strangelets in the atmosphere.
It should be mentioned that, for the sake of simplicity, our
calculation is limited to the ordinary (unpaired) strangelets,
i.e. the possible color-superconductivity effect of strangelets
has not been considered.

\section{Ground state properties of strangelets}
\label{}

It was argued that the phenomenal bag model first used by Alcock
\& Farhi~\cite{alcock1985} may not suit for strangelets.
Nevertheless, at the present level, bag model is still the most
effective way to understand the properties of strangelets, such as
ground state energy per baryon, charge-to-mass ratio,
fissionability, etc.

Within the framework of liquid drop model, He {\it et
al.}~\cite{he1996} studied ground state properties of strangelets
at finite temperature, by minimizing free energy of the system at
fixed baryon number. The Coulomb energy was neglected there
because the term contributes little to the system energy.
In this section, we just go one step further to include Coulomb
energy. Although Coulomb energy is negligible in computing $E/A$,
it can greatly affect $Z/A$, and hence the fissionability
parameters.

We consider strangelet as gas of u,d,s quarks, their antiquarks,
and gluons confined in an MIT bag model.
The grand potential of the system
$\Omega=\sum\limits_{i}\Omega_{i}+BV,$
where $B$ is the bag constant, $V$ is the volume, and the grand
potential of species $i$ is
\begin{equation}
\Omega_{i}=\mp
T\int_{0}^{\infty}dk\rho_{i}(k)\log(1\pm\exp(-(\sqrt{k^{2}+m_{i}^{2}}-\mu_{i})/T)).
\end{equation}
In the above equation, ``$\pm$'' denotes for Fermions/Bosons,
$\mu$ is the chemical potential, $\rho(k)$ denotes the density of
states, which is given by
\begin{equation}
\rho(k)=\frac{1}{2\pi^{2}}k^{2}V+f_{S}\left(\frac{m}{k}\right)kS+f_{C}\left(\frac{m}{k}\right)C,
\end{equation}
where $S$ ($=4\pi R^{2}$ for a sphere) is the surface area, $C$
($=8\pi R$ for a sphere) is the curvature.
The surface and curvature term for quarks are
$f_{S}^{(q)}(m/k)=-(1/8\pi)(1-(2/\pi)\arctan (k/m))$,
$f_{C}^{(q)}(m/k)=(1/(12\pi^{2}))(1-(3k/2m)(\pi/2-\arctan
(k/m)))$, respectively; for gluons, $f_{S}^{(g)}=0$,
$f_{C}^{(g)}=-1/(6\pi^{2})$.

The free energy of the system is given by
\begin{equation}
F=\sum_{i}{(\Omega_{i}+N_{i}\mu_{i})}+E_{coul}+BV,
\end{equation}
in which the term of Coulomb energy $E_{coul}=(3/5)\alpha Z^2/R$
if electric charge is uniformly distributed in the sphere, where
$\alpha$ is the fine structure constant and $Z$ is the total
electric charge.
Since Coulomb energy is taken into account, the chemical potential
of up quark and down quark (strange quark) are no longer
identical.

\begin{figure}[htb]
 \begin{center}
 \includegraphics{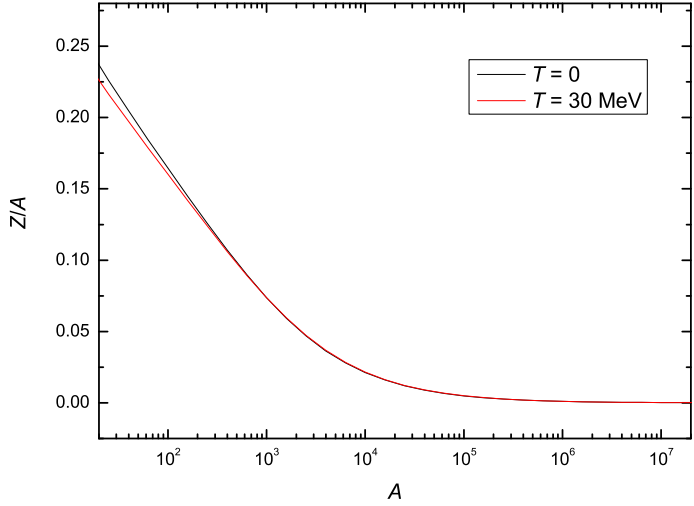}
 \caption{Charge to mass ratio of strangelets.}
 \label{fig1}
 \end{center}
 \end{figure}

By minimizing $F$, the chemical potentials, charge to mass ratio
and energy per baryon at any given baryon number and temperature
can be calculated.
Taking strange quark mass $m_{s}=150~\mathrm{MeV}$ and bag
constant $B=(145~\mathrm{MeV})^{4}$, we get the following fitting
values for the parameters. At zero temperature,
\begin{equation}
M_{str}/A = (314.6(4) A^{-0.532(4)} + 875.9(1)) ~\mathrm{MeV},
\end{equation}
therefore, the minimum baryon number for stability
($M/A<930~\mathrm{MeV}$) is $A_{crit}=27$.
The surface energy is
\begin{equation}
 E_{S}/A^{2/3} = (69.0(2) A^{-0.466(3)}+77.9(1)) ~\mathrm{MeV},
\end{equation}
and the rescaled radius is
\begin{equation}
r_{0} = R/A^{1/3}= (0.124(1) A^{-0.445(3)} + 0.941(1))
~\mathrm{fm},
\end{equation}
The numbers in parenthesis following each value indicate just the
fitting uncertainties of the value in the last digit.
The charge-to-mass ratio with respect to the equivalent baryon
number is shown in Fig.(\ref{fig1}).

As for a strangelet at excitation states to de-excite,
$\gamma-$ray emission, hadron emission and fission into small
parts should be under consideration.
$\gamma-$ray emission and meson emission do not change the baryon
number a little.
According to CEFT model~\cite{casher1979}, baryon evaporation is
suppressed in term of meson evaporation due to much smaller
probability to simultaneously form two quark-antiquark pairs than
one pair.
Banerjee {\it et al.}~\cite{banerjee1983} and Sumiyoshi {\it et
al.}~\cite{sumiyoshi1989} has calculated meson and baryon
evaporation rate of QGP ($\mu_{q}=0$), respectively. As for a
strangelet, in which the quarks have non-zero chemical potentials,
the numerical results are as follows.
The energy loss rate caused by baryon evaporation is
\begin{equation}
{dE \over dt}|_{baryon} = -6.23\times10^{19}A^{2/3}T^{2}{\exp
(-999.9/T)} ~\textrm{MeV}~\textrm{sec}^{-1},
\end{equation}
while the energy loss rate caused by meson evaporation is
\begin{equation}
{dE \over dt}|_{meson} = -1.12\times10^{20}A^{2/3}T^{2}{\exp
(-381.1/T)} ~\textrm{MeV}~\textrm{sec}^{-1}.
\end{equation}
Therefore, in the de-excitation process the only effective way to
change the baryon number of the strangelet is the fission.
Note that a strangelet at excited states will rapidly release its
energy in about $10^{-18}\sim10^{-15}$s, which is negligible
compared with colliding time intervals.

\section{Fission of strangelets by colliding with air nuclei}
\label{}

Now we investigate the stability of strangelets using the rotating
liquid drop model.
In the case of non-rotating systems, the relation between the
nature of the stationary points and the stability of a system is
simple, a maximum in one or more degrees of freedom indicates
instability.
However, the case for rotating systems is more subtle.

Consider a configuration of a rotating incompressible uniformly
charged fluid endowed with a surface tension.
The effective potential energy is given by
\begin{equation}
E=E_{S}+E_{C}+E_{R},
\end{equation}
where $E_{S}$ is the surface energy, $E_{C}$ the electrostatic
energy, and $E_{R}$ the rotational energy.
We neglect the curvature energy because it contributes little to
the issue we consider.

We may write the deformation energy, measured with respect to the
energy of the sphere, in the following dimensionless form,
familiar in the literature of nuclear fission,
\begin{equation}
\xi =
\frac{E-E^{(0)}}{E_{S}^{(0)}}=(B_{S}-1)+2x(B_{C}-1)+y(B_{R}-1).
\end{equation}
Here $B_{S}=E_{S}/E_{S}^{(0)}$, $B_{C}=E_{C}/E_{C}^{(0)}$,
$B_{R}=E_{R}/E_{R}^{(0)}$, and the two dimensionless parameters
$x$ and $y$ specify the ratios of electrostatic and rotational
energies of the sphere to the surface energy of the sphere, which
are defined as $x \equiv E_{C}^{(0)}/2E_{S}^{(0)}$ and $y \equiv
E_{R}^{(0)}/E_{S}^{(0)}$.
According to our calculations, we found that the fissionability
parameter $x$ of strangelets varies from 0.001 to 0.030, which is
much smaller than normal nuclei.
Therefore, we take $x=0$ for simplicity in the following
calculations.

If there is no rotation ($y=0$), the ground state is a sphere and
the saddle shape is the configuration of two tangent spheres.
With increased rotation, the ground state sphere is flatten into
an axially symmetric (Hiskes) shapes, and the saddle varies to the
so-called Pik-Pichak shapes. If $y$ is even larger, the ground
state will convert to a triaxial (Beringer-Knox) shape, which is
quite close in appearance to the Pik-Pichak saddle.
In fact, there exist a critical value $y_{crit}$ above which the
fission barrier vanishes, which implies a maximum rotation for
stability.

Cohen {\it et al.}~\cite{cohen1974} has calculated the ground
state and fission barrier energy measured with respect to the
energy of the sphere,
\begin{equation}
\xi_{ground} = y (-0.056 + 0.049y - 1.358y^{2} + 0.946y^{3}),
\end{equation}
\begin{equation}
\xi_{barrier} = 0.280 - 0.778y + 0.622y^{2} - 0.105y^{3}.
\end{equation}
For $x=0$, $y_{crit}=0.79$.

Now we consider a cosmic ray strangelet incident in the Earth's
atmosphere. The center-of-mass energy $E_{cm}=(M_{str}^{2} +
M_{air}^{2} + 2\gamma M_{str}M_{air})^{1/2}-M_{str}-M_{air}$,
where $M_{air}=14M_{0}$ ($M_0$: approximately the proton mass) is
the mass of the air nucleus, $M_{str}$ and $\gamma$ is the mass
and the Lorentz factor of the strangelet, respectively.
There exists a nonzero colliding cross section as long as the
strangelet has enough kinetic energy to overcome the Coulomb
barrier $E_{v}$ between the strangelet and the air nucleus, i.e.
if $E_{cm}>E_{v}$, they will have a chance to ``fusion'' into a
compound strangelet. In other words, the strangelet ``absorbs''
the air nucleus.

After fusion of the two, the excitation energy
$E_{e}=E_{cm}-E_{v}-E_{R}^{(0)}-E_{ground}$. According to rotating
liquid drop model, if the projectile strangelet is energetic
enough, $E_{e}$ will be higher than the fission barrier $E_{b}$,
the newly formed compound strangelet will fission into two smaller
strangelets which have nearly equal baryon numbers.

\begin{figure}[htb]
\begin{center}
\includegraphics{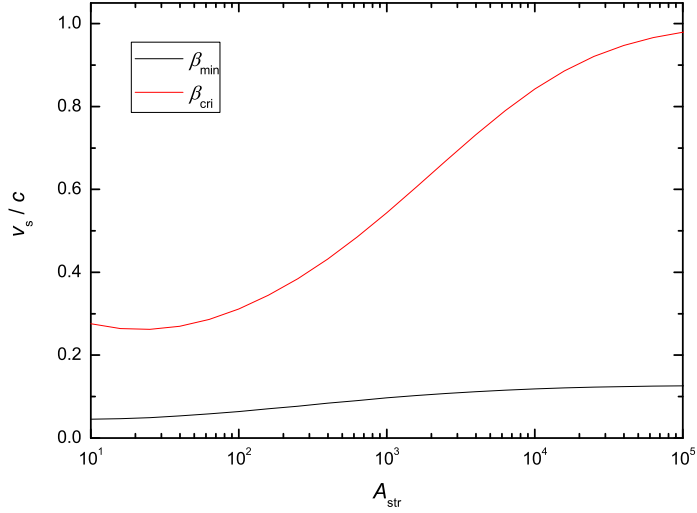}
\caption{Critical value of strangelet velocity. The upper curve
represents the fission threshold of a non-rotating strangelet, and
the lower curve represents the collision threshold.} \label{fig2}
\end{center}
\end{figure}

However, with the increasing kinetic energy of projectile
strangelet, the effect of rotation should no longer be neglected,
since if the fission barrier of rotating compound system
approaches zero, no compound strangelet will form.
It is reasonable to suppose the interaction time scale in this
case is much shorter than the former.

The geometric cross section for contact is $\sigma_{geo} = \pi
(r_{0}A_{str}^{1/3}+1.12A_{air}^{1/3})^{2}$, which is used by some
earlier studies for crude calculations.
However, the cross section is not always $\sigma_{geo}$ for
different baryon numbers and different velocities.
The cross section for close collision can be put as
\begin{equation}
\sigma_{col}=\sigma_{geo}
(1-Z_{str}Z_{air}e^{2}/(R_{str}+R_{air})E_{cm}),
\end{equation}
and the cross section for fusion is
\begin{equation}
\sigma_{cri} = 2\pi y_{crit}I_{0} E_{s}^{(0)}(M_{str}^{2} +
M_{air}^{2} + 2\gamma M_{str}M_{air})/((\gamma^{2} -
1)M_{str}^{2}M_{air}^{2}).
\end{equation}

\begin{figure}[htb]
\begin{center}
\includegraphics{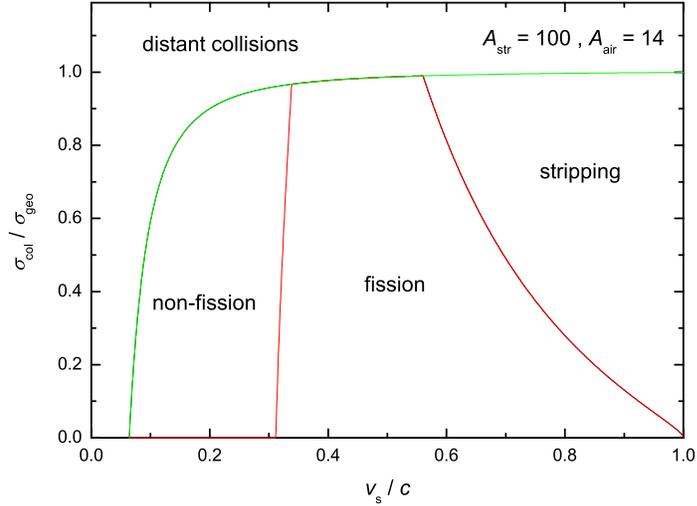}
\caption{A collision diagram of square of the impact parameter $b$
(times $\pi$) versus the velocity of a strangelet with
$A_{str}=100$, for the bombardment of an air nucleus by the
strangelet.} \label{fig3}
\end{center}
\end{figure}

Fig.(\ref{fig2}) gives the critical value of velocity for fusion
and fission as a function of baryon number.
The relation between the impact parameter $b$ in a collision
between the two masses and the velocity of the strangelet is shown
in Fig.(\ref{fig3}).

\section{Propagation of strangelets in the atmosphere}
\label{}

We consider strangelets with zero zenith angle of trajectory in
this work, and the influence of gravity force is neglected because
of its small contribution to the issue.
Therefore, strangelet-air collision and ionization effect are what
we mainly concern.

In our model, there exists a critical velocity of the strangelet
in the strangelet-air collisions below which the air nucleus will
be fused with the strangelet, and above which some mass will be
stripped from the strangelet.
The model is based on the analogical result in nuclear collisions,
i.e. the linear momentum transfer between the strangelet and the
air nucleus reaches a maximum around some critical energy.

When fusion happens, the velocity of the strangelet drops to
\begin{equation}
\gamma ~' = (\gamma M_{str} + M_{air})/(M_{str}^2 + M_{air}^2 +
2\gamma M_{str}M_{air})^{1/2}
\label{eq15}
\end{equation}
after each collision, and particularly if $E_{e}>E_{b}$, the
strangelet will fission into two smaller strangelets with equal
baryon number $0.5(A_{str}+A_{air})$ and probably equal
longitudinal velocity.

When the strangelet is more energetic, no compound strangelet will
form.
If the experimental law in nuclear collision is adaptable in this
case, the velocity of the strangelet is assumed to drop to
\begin{equation}
\gamma ~' = \gamma - (\gamma_{cri} - \gamma'_{cri}),
\end{equation}
where $\gamma_{cri}$ is the critical gamma factor for fusion, and
$\gamma'_{cri}$ can be deduced from Eq(\ref{eq15}) given
$\gamma=\gamma_{cri}$.
The value of $\gamma_{cri}$ can be found by solving the equation
$\sigma_{cri}=0.5~\sigma_{col}$, i.e. the watershed of
fusion-dominated and stripping-dominated collisions.
It should also be mentioned that, we assume new strangelet will
have a baryon number of $(A_{str}-A_{air})$ after each collision
in numerical calculations.
Indeed, at the present level, the mass and energy spectrum of
decay products in this range are quite uncertain.
Although our supposition is somewhat crude, it nevertheless tells
some important information.

In addition to the effect of colliding with air nuclei, the issue
of the energy loss of the strangelet through ionization of
surrounding media can not be avoided, which is described by the
Bethe-Bloch stopping power formula~\cite{sternheimer1984},
\begin{equation}
dE/dx=-0.153\beta^{-2}Z_{str}^{2}(\ln(\gamma^{2}-1)-\beta^{2}+9.39)~\textrm{MeV}/(\textrm{g~cm}^{-2})
\end{equation}
If $v<v_{0}Z_{str}^{2/3}$, where
$v_{0}=2.2\times10^{8}~\textrm{cm/s}$ is the speed of electron in
the first Bohr orbit, the effective charge $Z_{str} \rightarrow
(v/v_{0})Z_{str}^{1/3}$ due to the effect of electron
capture~\cite{bohr1948}.

The following figures show our numerical results.
In Fig.(\ref{fig4}), we present the mass evolvement of strangelets
with $\gamma_{0}=10^{3}$ as a function of atmospheric depth. It is
obvious that the larger the $\gamma_{0}$ and $A_{0}$, the more
possible the strangelet reach the sea level.
In Fig.(\ref{fig5}), we show the distribution of particular final
($x=1036~\textrm{g~cm}^{-2}$) baryon number of strangelets as a
function of initial baryon number and gamma factor. the upper
serried oblique lines correspond to products of fission (lower
energy), and the lower transverse lines correspond to products of
stripping (higher energy). We find that if the initial strangelets
have $A_{0}\geq3000$ or $\gamma_{0}\geq 140$, they will have a
chance to be detected by ground-based experiments.
In Fig.(\ref{fig6}), we give the final Lorentz factor as a
function of initial baryon number.

\begin{figure}[htb]
\begin{center}
\includegraphics{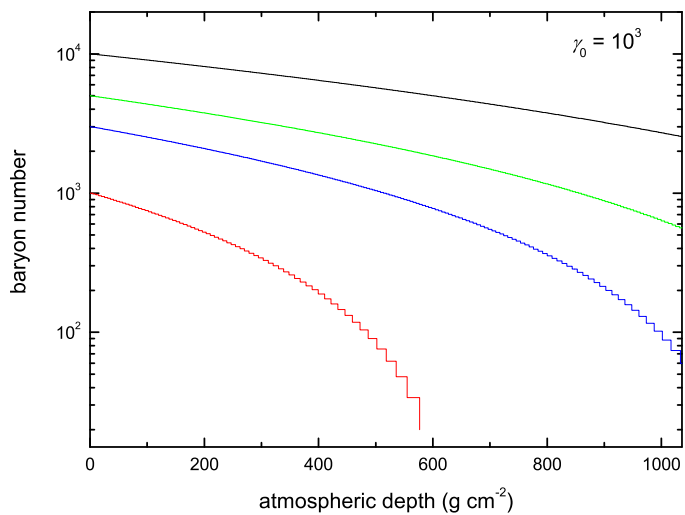}
\caption{Mass evolvement of strangelets with
$A_{0}=1000,3000,5000,10000$ and $\gamma_{0}=10^3$ as a function of
atmospheric depth.}
 \label{fig4}
\end{center}
\end{figure}

\begin{figure}[htb]
\begin{center}
\includegraphics{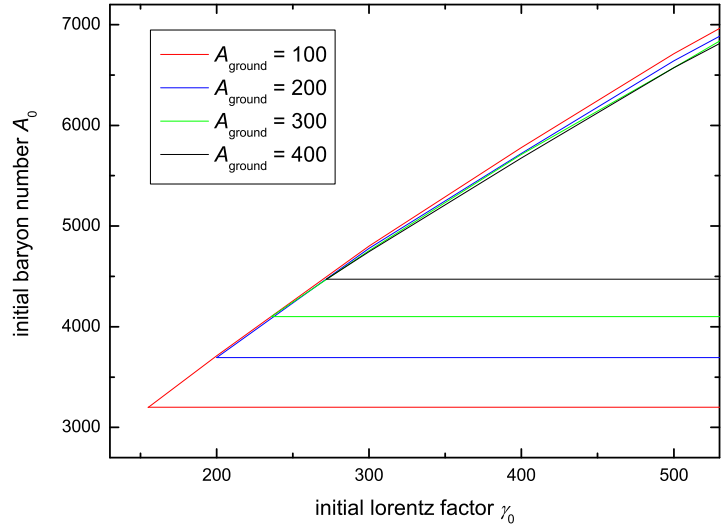}
\caption{Final mass distribution of strangelets
($x=1036~\textrm{g~cm}^{-2}$) as a function of initial baryon number
and lorentz factor.}
 \label{fig5}
\end{center}
\end{figure}

\begin{figure}[htb]
\begin{center}
\includegraphics{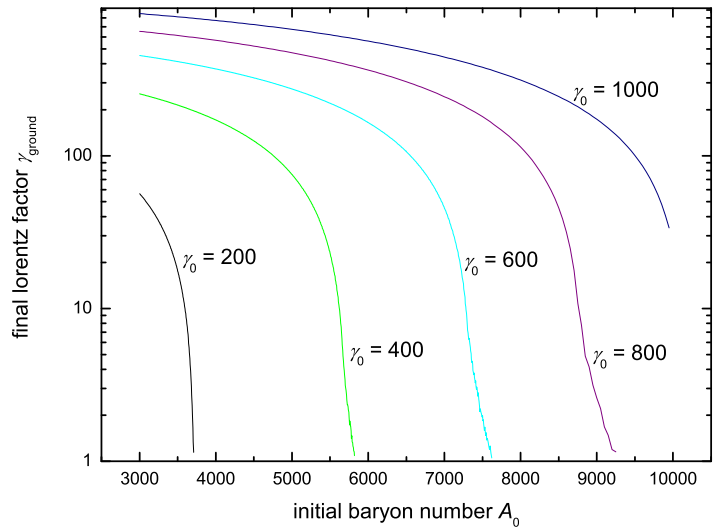}
\caption{Final lorentz factor ($x=1036~\textrm{g~cm}^{-2}$) as a
function of initial baryon number with
$\gamma_{0}=200,400,600,800,1000$.}
 \label{fig6}
\end{center}
\end{figure}

\section{Conclusions}
\label{}

From above discussion, it is reasonable to make a conclusion that
strangelets with large mass and energy have the chance to
penetrate the atmosphere to reach the sea level. Our model gives a
lower limit of initial baryon number, that is, $A_{0}\geq 3000$ or
$\gamma_{0}\geq 140$.
Relevant flux for ordinary strangelets is unclear yet, but as
predicted by Madsen~\cite{madsen2005}, it is about $1\sim10$ per
sqm year sterad, so there is a possibility for ground-based
experiments to detect them.
Madsen~\cite{madsen2005} predicts a flux of strangelets with a
velocity spectrum (an event with $\gamma = 140$ could be very
unlikely there) in a model where strangelets originate only by
merging of binary strange stars. However, the $\gamma$ factor of
strangelets produced in other ways (e.g., ejected and accelerated
in pulsar's magnetospheres~\cite{xw03,xu06,cu06}) may be greater
than 140, and the possibility of an event with higher $\gamma$
could then not be ruled out yet.

\begin{ack}
This work is supported by National Natural Science Foundation of
China (Nos.~10421503, 10575003, 10528510, 10573002), by the Key
Grant Project of Chinese Ministry of Education (No.~305001), and
by the Research Fund for the Doctoral Program of Higher Education
(China).
\end{ack}

\end{document}